\documentclass[slac_one]{revtex4}
\usepackage{graphicx}
\usepackage{fancyhdr}
\begin{document}

\title{{\small{2005 ALCPG \& ILC Workshops - Snowmass,
U.S.A.}}\\ 
\vspace{12pt}
Collider Signatures of SuperWIMP Warm Dark Matter} 

%

\author{Jose A.~R.~Cembranos}
\affiliation{Department of Physics and Astronomy,
 University of California, Irvine, CA 92697 USA}
\author{Jonathan L.~Feng}
\affiliation{Department of Physics and Astronomy,
 University of California, Irvine, CA 92697 USA}
\author{Arvind Rajaraman}
\affiliation{Department of Physics and Astronomy,
 University of California, Irvine, CA 92697 USA}
\author{Bryan T.~Smith}
\affiliation{Department of Physics and Astronomy,
 University of California, Irvine, CA 92697 USA}
\author{Fumihiro Takayama}
\affiliation{Institute for High-Energy Phenomenology,
Cornell University, Ithaca, NY 14853, USA}

\begin{abstract}
SuperWeakly-Interacting Massive Particles (superWIMPs) produced in the
late decays of other particles are well-motivated dark matter
candidates and may be favored over standard Weakly-Interacting Massive
Particles (WIMPs) by small scale structure observations. Among the
most promising frameworks that incorporate superWIMPs are $R$-parity
conserving supersymmetry models in which the lightest supersymmetric
particle (LSP) is the gravitino or the axino. In these well-defined
particle models, astrophysical observations have direct implications
for possible measurements at future colliders.
\end{abstract}
\maketitle

\thispagestyle{fancy}


\section{INTRODUCTION}

The dark matter (DM) problem is a one of the longstanding puzzles in
basic science.  Galaxy rotation curves, galaxy motions in clusters
and, more recently, precise measurements of the Cosmic Microwave
Background (CMB) radiation~\cite{CMBR}, Type Ia
supernovae~\cite{SNeIa}, large scale distribution of
galaxies~\cite{LSS} and Ly$\alpha$ clouds~\cite{Lyalpha} agree that
the Universe contains approximately five times more exotic matter than
ordinary. The standard candidates for DM are Weakly-Interacting
Massive Particles (WIMPs).  These emerge naturally in several
well-motivated particle physics frameworks, such as
supersymmetry~\cite{SUSY}, universal extra dimensions~\cite{UED} and
brane-worlds~\cite{BW1,BW2}. With masses and interactions at the
electro-weak scale, these are naturally produced with the correct DM
abundance. Furthermore, they behave as cold DM, which means that they
explain successfully the large scale structure of the Universe.

SuperWeakly-Interacting Massive Particles (SuperWIMPs) appear
naturally in the same well-motivated scenarios and their DM abundance
is also naturally the observed one, since they are produced in the
decays of WIMPs and naturally have similar masses. The cosmological
and astrophysical consequences of superWIMPs are, however, very
different from those of WIMPs. Consider, for example, minimal
supergravity (mSUGRA). In the WIMP scenario, the stau lightest
supersymmetric particle (LSP) region is excluded cosmologically.  In
the remaining region the neutralino is the LSP.  Much of the
neutralino LSP region is excluded because neutralinos are
overproduced, but some of this region is allowed.  In contrast, in the
superWIMP case, where, for instance, the gravitino or the axino is the
lightest supersymmetric particle (LSP), the region in which the stau
is the lightest standard model superpartner is very interesting, since
late decays of staus to gravitinos or axinos can impact Big Bang
nucleosynthesis (BBN) and possibly even resolve the $^7$Li
problem~\cite{Feng:2003xh}. At the same time, much of the
neutralino region is disfavored since neutralinos typically have
two-body decays that produce hadrons, which destroy BBN successes. In
fact, in the neutralino region, the region excluded by overproduction
in the WIMP scenario are the most interesting in the superWIMP one,
since the abundance of the dark matter is reduced by the ratio of WIMP
to superWIMP masses when the WIMPs decay.

\section{SUPERWIMP SIGNATURES}

SuperWIMPs signals are completely different from the WIMP
ones~\cite{Covi:1999ty, Covi:2001nw, Roszkowski:2004jd, Ellis:2003dn,
Feng:2003xh, Feng:2003nr, Feng:2004zu}. As noted above, late decays to
superWIMPs have implications for BBN.  At the same time, late decays
can also distort the Planckian spectrum of the Cosmic Microwave
Background, which means that new experiments like ARCADE and DIMES can
find evidence for these particles. 

\begin{figure}[bt]
\centerline{
\includegraphics[
width=8cm,height=6.7cm,clip]{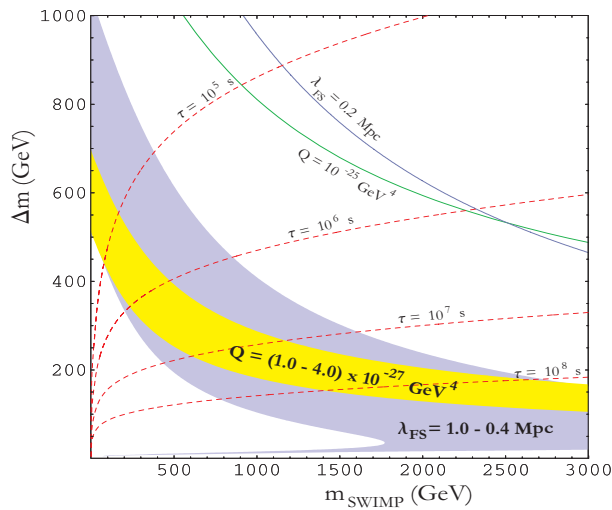}} 
\caption{Regions (shaded) of the $(m_{\text{SWIMP}}, \Delta m)$ plane
 preferred by small scale structure observations, where $\Delta m
 \equiv m_{\text{NLSP}} - m_{\text{SWIMP}}$, for gravitino superWIMPs
 with sneutrino NLSPs.  The regions under both bands are disfavored.
 In the regions above both bands, superWIMP DM becomes similar to cold
 DM.  Contours of typical lifetimes $\tau_{\tilde{\nu}}$ are also
 shown~\cite{Cembranos:2005us}. }
\label{sneu}
\end{figure}

SuperWIMP scenarios also differ from WIMP scenarios in their
implications for structure formation in the Universe.  Because
superWIMPs are produced with large velocities in late decays, they can
behave as warm DM and resolve some apparent discrepancies of
observations with cold DM. Comparing some simulations with
observations, WIMPs predict overdense cores in galactic halos, one or
two orders of magnitude more dwarf galaxies in the Local Group than
observed, and disk galaxies with less angular momentum.  The velocity
and angular momentum of DM halos can be increased naturally in
superWIMP scenarios.  As shown in Figure \ref{sneu} from
Ref.~\cite{Cembranos:2005us}, this has been supported by analyses of
phase space densities and studies of damping in the power
spectrum~\cite{Cembranos:2005us,Kaplinghat:2005sy,Sigurdson:2003vy}.

The collider phenomenology of superWIMPs have been analyzed from
different points of
view~\cite{Buchmuller:2004rq,Feng:2004gn,Hamaguchi:2004df,Feng:2004yi,Brandenburg:2005he,Feng:2005gj}.
We have studied the small scale structure consequences of pure
superWIMP DM scenarios.  These analyses show that superWIMP signals
can be observed even in the ILC not only for the warm DM case,
but also of the cold DM one.

Because superWIMP scenarios are most naturally accompanied by charged
particles that decay after seconds to months, it has been proposed
that these charged particles can be trapped outside of the particle
physics detector so that their decays can be studied.  A liquid trap
can be used to stop the charged particles so they can be transported
to a quiet environment~\cite{Feng:2004yi}, and an active stationary
detector/trap has also been proposed~\cite{Hamaguchi:2004df}.  The
liquid trap will cost less and allow for transportation to a clean,
low background environment but will be incapable of measuring the
shorter range of the lifetimes.  The active detector can measure
almost the entire range of the lifetimes, but must differentiate late
decays from backgrounds from the high energy experiment and cosmic
rays.  Both methods of trapping can be optimized by choosing the
correct shape and placement of the traps to catch the maximum number
of the meta-stable charged particles.  By trapping many charged
particles and studying their decays, the properties of the decay
products may be indirectly constrained, providing an accurate
identification of the LSP and superWIMP DM.

\section{CONCLUSIONS}

We have examined the implications of superWIMPs, whose behaviour is,
in general, very different to WIMPs.  Analyses of astrophysical data
constrain superWIMP scenarios, and also motivate some spectacular
possibilities for collider signatires of new physics.

If DM is composed of superWIMPs, charged slepton NLSPs, or similar
particles, will appear stable because of the length of its lifetime. This
superWIMP signature could be studied at both the LHC and the ILC.

\begin{acknowledgments}
JARC acknowledges the hospitality and collaboration of workshop
organizers and conveners, and support from the NSF and Fulbright OLP. 
The work of JARC is supported in part by NSF grant
No.~PHY--0239817, the Fulbright-MEC program, and the FPA 2005-02327
project (DGICYT, Spain). The work of JLF is supported in part by NSF
grant No.~PHY--0239817, NASA grant No.~NNG05GG44G, and the Alfred
P.~Sloan Foundation. The work of AR is supported in part by NSF grant
No.~PHY--0354993.  The work of FT is supported in part by NSF grant
No.~PHY--0239817.
\end{acknowledgments}

\end{document}